\documentclass[aps,pra,numerical,reprint,showpacs]{revtex4-1}
\usepackage[dvips]{graphicx}%
\usepackage{bm}%
\usepackage{amsmath}
\usepackage{dcolumn}
\usepackage[colorlinks=true,linkcolor=blue,citecolor=red ,urlcolor=blue]{hyperref}

\begin{document}

\title{Multiscale Modeling of the effect of Pressure on the Interfacial Tension and other Cohesion Parameters in Binary Mixtures}
\author{E. Mayoral$^\dagger$ and E. Nahmad-Achar$^\ddagger$}

\maketitle
$^\dagger$
Instituto Nacional de Investigaciones Nucleares, Carretera M\'exico-Toluca S/N, La Marquesa Ocoyoacac, Edo. de M\'exico
C.P. 52750, M\'exico\\
$^\ddagger$
Instituto de Ciencias Nucleares, Universidad Nacional Aut\'onoma de M\'exico, Apartado Postal 70-543, 04510 M\'exico DF,   Mexico  \\


{\it pacs 42.50.Ct, 42.50.Nn, 73.43.Nq, 03.65.Fd}


\section{Abstract}
We study and predict the interfacial tension, solubility parameters and Flory-Huggins parameters of binary mixtures as functions of pressure and temperature, using multiscale numerical simulation. A mesoscopic approach is proposed for simulating the pressure dependence of the interfacial tension for binary mixtures, at different temperatures, using classical Dissipative Particle Dynamics (DPD). The thermodynamic properties of real systems are reproduced via the parametrization of the repulsive interaction parameters as functions of pressure and temperature via Molecular Dynamics simulations. Using this methodology, we calculate and analyze the cohesive density energy and the solubility parameters of different species obtaining excellent agreement with reported experimental behavior. The pressure - and temperature-dependent Flory-Huggins and repulsive DPD interaction parameters for binary mixtures are also obtained and validated against experimental data. This multiscale methodology offers the benefit of being applicable for any species and under difficult or non-feasible experimental conditions, at a relatively low computational cost.

\section{Introduction}

Pressure is a fundamental thermodynamic variable in many industrial processes. The study, understanding and prediction of how changes in pressure affect different properties such as solubility parameters, activity coefficients, Flory-Huggins parameters and interfacial tension, amongst others, is fundamental in the design and application of multi-component products. In particular, the understanding of hydrocarbon-water mixtures at different thermodynamic conditions is essential in the process of oil recovery and other industrial applications. In these systems the effect of high pressure and temperature in different properties is particularly relevant in order, for example, to improve the displacement kinetics involved during the extraction of oil. Complex capillary responses due to changes in interfacial tension and in cohesion parameters originated by variable thermodynamic conditions in the oil reservoir could have important economic consequences. The  evaluation of these effects is not easy to perform in the laboratory due to the fact that experiments in such extreme conditions are expensive and difficult to handle. For this reason there are many measurements under atmospheric conditions, but only a few studies at high pressure and temperature are available\cite{experimentaltension1,experimentaltension2}.

Interfacial tension $\sigma$ and cohesion parameters such as the cohesive energy density ${\cal E}_{coh}$, the  solubility $\delta$ and the Flory-Huggins $\chi$ parameters are important quantities widely used in different industrial areas such as paints and coatings, pharmaceutics, bio-polymers, membranes, smart materials, etc. A paramount goal in this area is to establish an accurate and accessible methodology to obtain these parameters for complex mixtures as a function of pressure and temperature. Usually, the solubility parameters are obtained experimentally by means of the heat of vaporization  at atmospheric conditions, but in a complex fluid system the components are not totally volatile at different thermodynamic conditions \cite{RefVerdier}. Other alternatives such as the use of equations of state could be employed but in this case it is necessary to have a good description of the volume and the density behavior at different temperatures and pressures which is not an easy task. Another common solution has been to consider the internal pressure as  a substitute for the solubility parameters \cite{RefHildebrand, RefBarton} but these two concepts do not describe the same phenomena especially under different thermodynamic conditions \cite{RefVerdier}.

 As an alternative for the study of this kind of complex systems, multiscale numerical simulation has shown to be a very promissing option \cite{RefMurtola}. As different time and length scales and a big number of components are involved in complex fluids, multiscale modeling involving atomistic and mesoscopic approaches has been considered as an attractive combination for their study. The use of these two techniques permits to simulate large complex systems taking advantage of the collective properties at a mesoscopic level with relatively cheap computational requirements, and also to calculate properties at the atomistic level when necessary.

 As it is well known, the collective and cooperative behavior emerge in systems with many particles making possible the use of coarse grained simulations and scaling concepts for their study. One of these numerical techniques is the Dissipative Particle Dynamic (DPD) methodology \cite{RefHoggenber} which is specially attractive to simulate correctly the hydrodynamics of complex liquids (for an extensive review of this methodology see for example \cite{RefMurtola, RefEMVENASpringer}).

Even though the use of classical DPD has been applied to simulate different kinds of systems, its use to model and reproduce real behavior in a DPD fluid at different thermodynamic conditions (different pressures or temperatures), not only in a qualitative but also in a quantitatively way, remains a challenge due to the restricted thermodynamic behavior given by the functional expressions for the conservative force employed .

Some efforts have been done in order to describe the thermodynamic properties of real systems with DPD using modifications  of the traditional technique through, for example, the so called Multibody-DPD (MDPD)\cite{RefTrofimov14, RefTrofimov15}. In MDPD the conservative force depends not only on the inter-particle separation but also on the instantaneous local particle density which depends on the positions of all other neighboring particles. For this reason, the conservative force in MDPD is a many-body force. Using an improved MDPD model some authors \cite{RefTrofimov,RefJakovsen}, have performed simulations for single and multicomponent systems at constant pressure, including a modified version of the Andersen barostat to suppress the unphysical volume oscillations due to pressure changes, simplifying the equilibration of the system.

Nevertheless, the effect of pressure using the classical DPD technique has remained unexplored. In traditional DPD the thermodynamic quantities are obtained using only a single parameter $a_{ij}$ in the conservative force. Recently, one of us has presented the methodology to model the effect of temperature in this kind of simulations via a temperature-dependent repulsive $a_{ij}(T)$ parameter \cite{RefEMVAGGT1, RefEMVAGGT2}. This parametrization allows one to consider the thermodynamic conditions  using information at an atomistic level. 

In this contribution we present the technique to study the effect of pressure with the DPD approach, following a similar idea.  To achieve this, we obtain and analyze first the effect of pressure and temperature on the cohesive energy density ${\cal E}_{coh}(T,P)$ for the pure components via atomistic simulations, and then calculate the solubility parameters $\delta(T,P)$ and the Flory-Huggins $\chi_{ij}(T,P)$ parameters in binary liquid-liquid mixtures at different pressures and temperatures. Following the standard DPD methodology, we assume that the temperature and pressure dependence of the DPD $a_{ij}(T,P)$ parameters could be associated directly with $\chi_{ij}(T,P)$.

We probe this direct dependence by modeling the interfacial tension between binary mixtures performing DPD simulations. We use our model to predict the interfacial tension of benzene-water and n-decane-water mixtures at T = 298, 323 and 373 K and P = 200, 400, 600, 700, 800, 1000 and 1200 atmospheres, and compare with reported  experimental data \cite{experimentaltension1,experimentaltension2} obtaining an excellent agreement. To our knowledge, this is the first time that DPD simulations at different pressures and temperatures are reported with such exactitude.

In Section II we describe the general characteristics of DPD methodology and the procedure followed to obtain the pressure dependence of the DPD interaction parameters via the solubility and Flory-Huggins parameters. In Section III we discuss the effect of pressure and temperature in the solubility parameters and interfacial tension. Section IV presents the simulation details and the results obtained for the cohesive density energy ${\cal E}_{coh}$ and the solubility parameters $\delta$ by atomistic simulations at several pressures and temperatures, as well as for the interfacial tension $\sigma$ of benzene-water and n-decane-water mixtures performing DPD coarse grained simulations. Finally, some conclusions are discussed in Section V.

\section{Modeling the effect of pressure with DPD}

The core structure of the Dissipative Particle Dynamics (DPD) simulation method \cite{RefHoggenber,RefGroot} is fundamentally the same as a classic Molecular Dynamics (MD) algorithm but, in DPD, the particles correspond to coarse–grained structures representing molecular or atomic clusters instead of individual atoms. The momentum and position of each DPD particle are calculated by solving Newton's second law of motion using the total force acting on it at finite time steps. The main difference involving MD and DPD methodology is that, in DPD, the functional structure of the interacting force linking any two particles $i$ and $j$ is constituted by the sum of three components: a conservative ($\bm{F}_{ij}^C$), a dissipative ($\bm{F}_{ij}^D$), and a random ($\bm{F}_{ij}^R$) force. The total  force felt by particle $i$ due to the presence of all other particles is thus
\begin{equation}\label{FuerzasDPD}
\bm{F}_{i}=\Sigma_{j\neq i}^N [\bm{F}_{ij}^C + \bm{F}_{ij}^D + \bm{F}_{ij}^R]
\end{equation}

The time evolution of velocities and positions are calculated from $\bm{\dot v_i} = \bm{F}_i$ and  $\bm{\dot r_i}  = \bm{v_i}$ , where $\bm{v_i}$ is the velocity and  $\bm{r_i}$ the position of particle $i$.
A soft, linearly decaying repulsive interaction is used for the conservative force between each particle pair:

\begin{equation}
	\bm{F}_{ij}^C =
	\begin{cases}
	a_{ij}\,(1-r_{ij}/r_c)\,\hat{\bm{r}}_{ij}, &\text{$(r_{ij} < r_c)$} \\
	0, &\text{$(r_{ij} \geq r_c)$}.
	 \end{cases}
\end{equation}

\noindent where $\bm{r}_{ij} = \bm{r}_i - \bm{r}_j, r_{ij} = |\bm{r}_{ij}|$ and $\hat {\bm{r}} ̂_{ij} = \bm{r}_{ij}/r_{ij}$. In this expression $a_{ij}$ is known as the repulsive DPD parameter acting between the pair of particles and $r_c$ represents a cutoff distance. The dissipative and the random forces are defined as:
\begin{subequations}
\begin{eqnarray}
\bm{F}_{ij}^D &=&  -\gamma\omega^D(r_{ij})[\bm{\hat{r}}_{ij}\cdot\bm{v}_{ij}]\hat {\bm{r}}_{ij} \,,\\
\bm{F}_{ij}^R &=&  \sigma\omega^R(r_{ij})\xi_{ij}\hat{\bm{r}}_{ij}\,,
\end{eqnarray}
\end{subequations}

\noindent respectively. Here, $\sigma$ is the noise amplitude and $\gamma$ is the friction coefficient. To guarantee that a Boltzmann distribution is achieved at equilibrium, $\sigma$ and $\gamma$ are related by $k_B T=\frac{\sigma^2}{2\gamma}$  as a consequence of the fluctuation - dissipation theorem \cite{RefEspanol}, keeping the temperature internally fixed. Here $k_B$ is Boltzmann's constant, $\xi_{ij} = \xi_{ji}$ is a random number distributed between $0$ and $1$ with Gaussian distribution and unit variance, and $\bm{v}_{ij} = \bm{v}_i - \bm{v}_j$ is the relative velocity between the particles. The weight functions $\omega^D$ and $\omega^R$ depend on distance and vanish for $r \geq r_c$  , they are commonly established as:

\begin{equation}
\label{omegas}
\omega^D(r_{ij})=[\omega^R(r_{ij})]^2 =
\begin{cases}
	(1-r_{ij}/r_c)^2, &\text{$(r_{ij} \leq r_c)$} \\
	0, &\text{$(r_{ij} > r_c)$}.
	 \end{cases}
\end{equation}

The cutoff radius $r_c$ is commonly chosen as the reduced unit of length, $r_c = 1$, and corresponds to the intrinsic length scale of the DPD model. The masses of all particles are chosen to be equal to the reduced unit of mass $m_i = 1$. The cut-off radius $r_c$ could be written as:

\begin{equation}
r_c= (v_b N_m \rho^* )^{1/3}
\label{rc}
\end{equation}
where $v_b$ is the bead volume, $N_m$ is the coarse graining factor (it is the number of water molecules per DPD particle) and can be considered as a real-space renormalization factor, and  $\rho^*$ is the number of DPD beads in a lattice box of side $r_c$ and volume $r_c^3$. For example, for $N_m = 1$,  $\rho^*=3$, and $v_b=18$\AA${}^3$, we have $r_c = 4.48$\AA. 

As we have mentioned, thermodynamic quantities in DPD are calculated using only the conservative force \cite{RefFrenkel}, and for this reason a correct estimation and scaling for the repulsive $a_{ij}$ parameter is fundamental to simulate realistic systems at different thermodynamic conditions. The effect of changes in the random and dissipative forces has been recently studied \cite{RefAGGgammasigma} showing that the average contribution of these two forces to the pressure is negligible. For this reason we will focus only on the conservative parameter $a_{ij}$.
The DPD parametrization needs to be appropriately selected to give an accurate estimate of the real system. For sufficiently large number densities the  DPD equation of state for a monocomponent system is~\cite{RefGroot, RefPivkin}

\begin{equation*}
	P = \rho\,k_BT + \alpha\,a_{ii}\rho^2\, ,
\end{equation*}

where $ \alpha = 0.101\pm 0.001$ is a numerical constant and $a_{ii}$ are the interaction parameters for identical DPD particles.  
This equation shows the dependence on $P$ of the interaction parameters. To describe the real system, fluctuations in the liquid must be described adequately, and these are determined by the compressibility of the system. The  definition for the dimensionless isothermal compressibility is $\kappa^{-1} = 1/nk_BT\kappa_T=1/k_B T(\partial P/\partial n)_T$ where $n=N_m\,\rho_{DPD}$  is the number density of molecules in the medium and $\kappa_T$ is the usual isothermal compressibility, $(\kappa_T)^{-1} = n\,(\partial P/\partial n)_T$. The dimensionless isothermal compressibility for water at standard conditions is $\kappa^{-1} = 15.9835 \approx 16$, and may be considered to be constant in the pressure range from $20$ to $120$ MPa to be considered~\cite{RefWilhem}. 
Using the DPD equation of state the following relationship emerges:

\begin{eqnarray}
\!\!\!\!\!\!\!\!\!\!\ \kappa^{-1}&=&\frac{1}{k_B T}(\frac{\partial P}{\partial\, \rho_{DPD}})_T\,(\frac{\partial\, \rho_{DPD}}{\partial n})_T\\ \nonumber
&=& \frac{1}{N_m}(1+2\alpha \frac{a_{ii}\,\rho_{DPD}}{k_BT}).
\end{eqnarray}

Then, the conservative force parameter for particles of the same type, $a_{ii}$, may be obtained as

\begin{equation}
a_{ii}=[\frac{\kappa^{-1} N_m-1}{2\,\alpha\, \rho_{DPD}^* }] k_B T.
\label{aii}\end{equation}
These equations give the relationship between the mesoscopic model parameter and the real compressibility of the system.
The free energy density for a DPD monocomponent system is

\begin{equation}
\frac{f_v}{k_BT}= \rho \ln\rho -\rho + \frac{\alpha\, a_{ii}\,\rho^2}{k_B T}
\end{equation}
and for a two-component system
\begin{eqnarray}
\!\!\!\!\!\!\!\!\!\!\ \frac{f_v}{k_BT}=\frac{\rho_i}{N_i}\ln \rho_i + \frac{\rho_j}{N_j}\ln \rho_j -\frac{\rho_i}{N_i}-\frac{\rho_j}{N_j} \\ \nonumber
+ \frac{\alpha(a_{ii}\rho_i^2+2a_{ij}\rho_i\rho_j+a_{jj}\rho_j^2)}{k_BT}
\end{eqnarray}
where the indices $i,\,j$ refer to species $i$ and $j$ respectively.
Writing $a_{ii}=a_{jj}$, $\rho_i+\rho_j = \rho$ and $x = \rho_i/(\rho_i+\rho_j)$, we have

\begin{equation}
\frac{f_v}{\rho k_BT}\sim \frac{x}{N_A}\ln x + \frac{(1-x)}{N_B}\ln(1-x) + \chi x (1-x) + const.
\end{equation}
where
\begin{equation}
 \chi = \frac{2 \alpha (a_{ij}-a_{ii})\rho}{k_BT}
\end{equation}
has been identified with the well known Flory-Huggins parameter. Groot and Warren~\cite{RefGroot} found that there is a linear relation between $\chi$ and $\Delta a = a_{ii}-a_{ij}$ given by $\chi= (0.286\pm 0.002)\Delta a $ for $(\rho=3)$.

We propose that the dependence on temperature and pressure could be then included in the DPD repulsive parameter via the pressure and temperature dependent Flory-Huggins parameter $\chi_{ij}(T,P)$, and can be generalized as

\begin{equation}
a_{ij}(T,P) = a_{ii}+ \frac{1}{0.286}\chi_{ij}(T,P)
\label{aijPT}
\end{equation}
where ${0.286}^{-1}$ is a numerical constant \cite{RefGroot} and  $\chi_{ij}(T,P)$  is given by
\begin{equation}
\chi_{ij}(T,P)=\frac{V_i (T,P)}{RT}[\delta_i (T,P)-\delta_j (T,P)]^2.
\label{chiPT}
\end{equation}
Here, $\delta_i (T,P)$ is the solubility parameter of the $i$-th  particle, which we take to be that of the species it represents even if the particle does not cover a full molecule of it (\emph{vide infra} for simulation details),  and $V_i(T,P)$ is the partial molar volume of particle $i$ at temperature $T$ and pressure $P$.

Interfacial tensions are obtained by the Irving-Kirkwood~\cite{IK} method expressing the surface tension from the local components of the pressure tensor. We use the virial theorem route \cite{RefAllen} and the components of the pressure tensor $P_{ii}(i = x, y, z)$ obtained from the total conservative force, and time averages over the simulation time. For an interface at the x-y plane we have

\begin{eqnarray}
\!\!\!\!\!\!\!\!\!\!\sigma(T,P) = &&\int_{-L_z/2}^{L_z/2} \{ \langle P_{zz}(T,P;\,z)\rangle\nonumber\\
 &&- \frac{1}{2}[\langle P_{xx}(T,P;\,z) \rangle + \langle P_{yy}(T,P;\,z)\rangle]\}dz
\label{tensionsim}
\end{eqnarray}
were $L_z$ is the length of the simulation box in the z-direction, the brackets indicate time average over the integration phase of the simulation,  and $P_{ii}(T,P)$ are the temperature and pressure dependent component of the pressure tensor in the $i$-direction; similarly for other interfaces.

\section{Effect of pressure and temperature in the interfacial tension and in the cohesion parameters}

It is known that the effect of temperature in the interfacial tension is much greater than that of the pressure. Usually, the maximum change in the interfacial tension in binary mixtures as a function of pressure, in the range $20$ to $120$ MPa, is around $0.5$ to $2$ dynes/cm; for this reason the effect of pressure is difficult to observe in the laboratory. Experimental studies using binary mixtures (benzene/water and n-decane/water) have suggested the following phenomenological relation over a range of 20 to 150 $^\circ$C and for pressures of 200 to 700 Atm \cite{experimentaltension2}:

\begin{equation}
\sigma = a_0 + a_1 P + a_2 \Delta T
\label{experimentaltension}
\end{equation}
where $P$ is the pressure in atmospheres and $\Delta T = T - 25$  in  $ ^{\circ}$C. The values of the coefficients $a_0$, $a_1$ and $a_2$ depend on the kind of liquids in the system.

Some authors \cite{experimentaltension2, experimentaltension3} have reported that the pressure coefficient $a_1$ is very small and positive for the case of benzene/water and n-decane/water systems for pressures lower than 700 atmospheres, but experimental results obtained in \cite{experimentaltension1} for the benzene/water system at pressures higher than 700 atmospheres show a negative pressure coefficient as a consequence of the increased influence in the solubility of the phases when the pressure is increased. The decrease in the interfacial tension when the pressure is increased has been reported also for binary mixtures of CO$_2$/alkane systems \cite{experimentalbueno3, experimentalbueno1}.

 The forces involved in the interfacial tension of multicomponent fluids are intimately related to the solubility and Flory-Huggins parameters. The concept of the solubility parameter $\delta$  introduced by Hildebrand and Scott in 1950 \cite{RefHildebrand} has been fundamental in the analysis of mixtures and pure compounds in many industrial areas. It is usually assumed that the solubility parameter consists of a linear combination of contributions  from dispersion interactions, polar interactions and hydrogen bonding \cite{RefHansen}:
\begin{equation}
\delta^2 = \delta^2_{d} + \delta^2_{p} + \delta^2_{h}
\end{equation}

Commonly, this parameter is calculated at atmospheric pressure via the heat of vaporization, but in many cases high pressures and temperatures are involved in industrial processes and this procedure could lead to a poor estimate for $\delta$. A more adequate estimation of $\delta(T,P)$ under different conditions of pressure and temperature is necessary, and we can express it in terms of the cohesive energy of the liquid  $E_{coh}(T,P)$ and its molar volume $V(T,P)$ as
\begin{equation}
	\label{solubilityparam}
\delta(T,P)=\left(\frac{-E_{coh}(T,P)}{V(T,P)}\right)^{1/2}= {\cal E}_{coh}(T,P)^{1/2},
\end{equation}
where the cohesive energy density (${\cal E}_{coh}$) is a measure of the whole molecular cohesion per unit volume.
 The effect of pressure and temperature on the solubility parameter has been estimated by Null and Palmer \cite{RefNull}  for vapor pressure calculations as
\begin{equation}
\delta(T,P)=\left[\frac{2.303R B}{V_{liq} (T,P)} \left(\frac{T}{T + C -273.15}\right)^2 - \frac{RT}{V_{liq} (T,P)}\right]^{1/2}
\label{deltanull}
\end{equation}
where $V_{liq}(T,P)$ is the molar volume of the liquid phase, $R$ the gas constant and $B$, $C$ are the constants of Antoine's equation \cite{RefAntoine}
\begin{equation}\label{antoine}
\log⁡ P =A - \frac{B}{C+T}
\end{equation}
with $P$ and $T$  the pressure and temperature of the vapor phase. Antoine's equation permits a good estimation of the vapor pressure as a function of temperature, but unfortunately the constants are not available for all systems.Those for the species considered in this work are shown in Table~\ref{table0}.

 For the dependence of the cohesive energy on pressure and temperature we use Barton's expression \cite{RefVerdier,RefBarton}
\begin{equation}
	\label{ecoh}
- {E}_{coh}(T,P) = U_{vap}(T,P = 0) - U_{liq}(T,P) = - U_r
\end{equation}
where $U_{vap}$ and $U_{liq}$ are the internal energy of the vapor and liquid phases respectively \cite{RefVerdier}, and $U_r$ is the residual internal energy. MacDonald and Hyne \cite{RefMacDonald} determined the cohesive energy density of alcohol-water mixtures at different temperatures and at atmospheric pressure finding that the cohesive energy density has a monotonous behavior with P over a very large range.

In this contribution we present an alternative to obtain all this information via multi-scale simulations using Molecular Dynamics and Dissipative Particle Dynamics simulations and compare the results with the experimental data.

\section{METHODOLOGY AND RESULTS}

\subsection{Solubility parameters as a function of pressure and temperature by molecular dynamic simulation}
\label{sec4A}
The solubility parameters $\delta(T,P)$ and cohesive energy densities ${\cal E}_{coh}(T,P)$ at temperatures $T = 298, 323, 373 K$ and pressures $P = 200, 400, 600, 700, 800, 900, 1000, 1200$ Atm for benzene, n-decane and water were calculated performing atomistic molecular dynamics simulations. We consider periodic cells of amorphous fluid structures, using the Amorphous Cell module of the Materials Studio suite \cite{RefMS}. The dimension of the simulation box was chosen in all cases to be $25$ $\textup{\r{A}}$ and the COMPASS force field was used to model the interatomic interactions. We developed $NPT$ dynamics simulations in order to equilibrate the density of the system at the temperature and pressure of interest. We then used the Discover Molecular Dynamics engine to evolve the systems at these thermodynamic conditions obtaining statistically independent structures.

From this information we obtained the total solubility parameter $\delta$[J/cm$^3$] as well as the electrostatic $\delta_e$[J/cm$^3$] and dispersive $\delta_{VdW}$[J/cm$^3$] contributions for each component, the cohesive energy density ${\cal E}_{coh}$[J/m], the density $\rho $[gr/cm$^3$],  and the molar volume $V_m$[$\textup{\r{A}}^3$].

\begin{figure*}[h!]
\scalebox{0.25}{\includegraphics{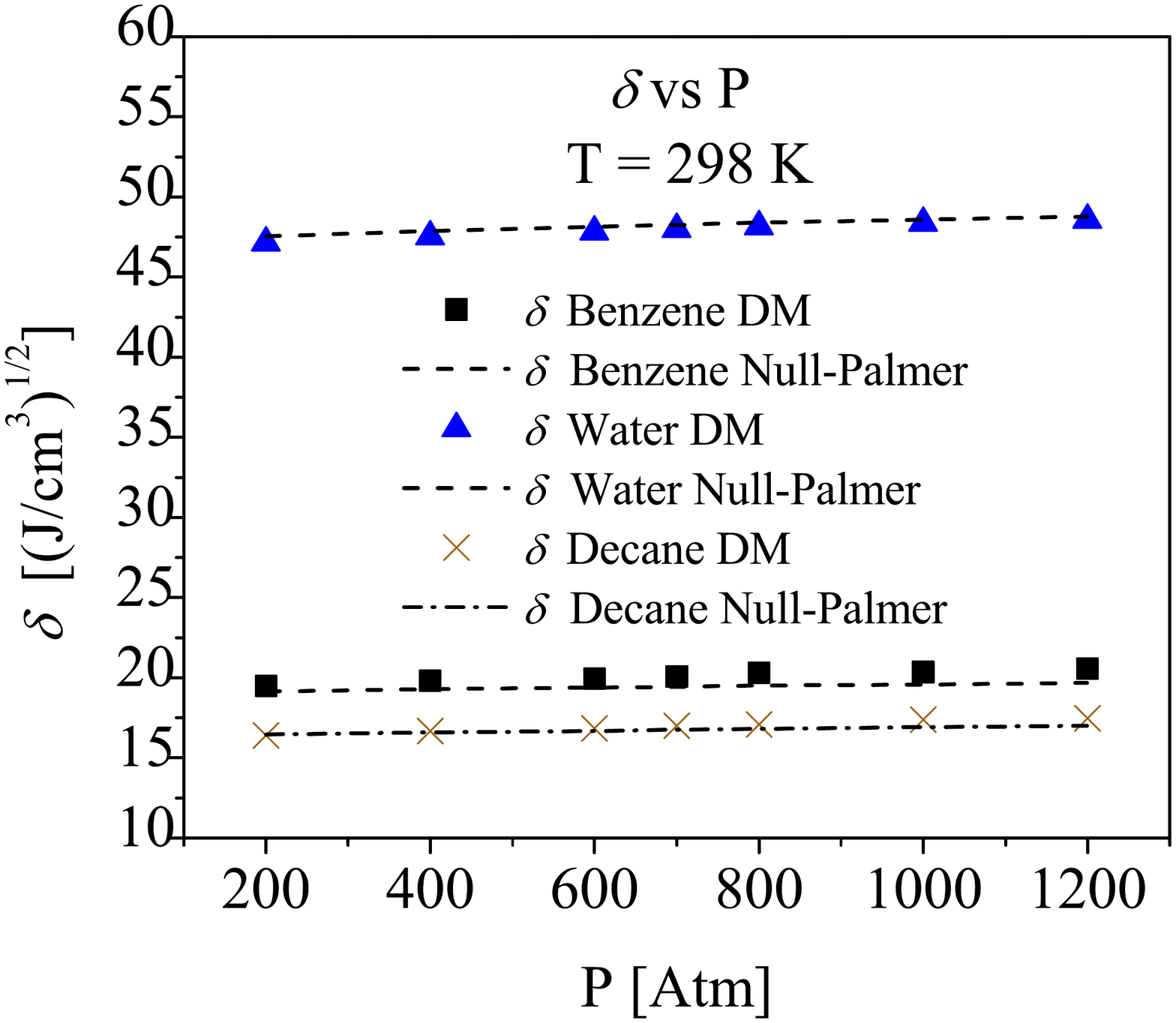}}\quad
\scalebox{0.25}{\includegraphics{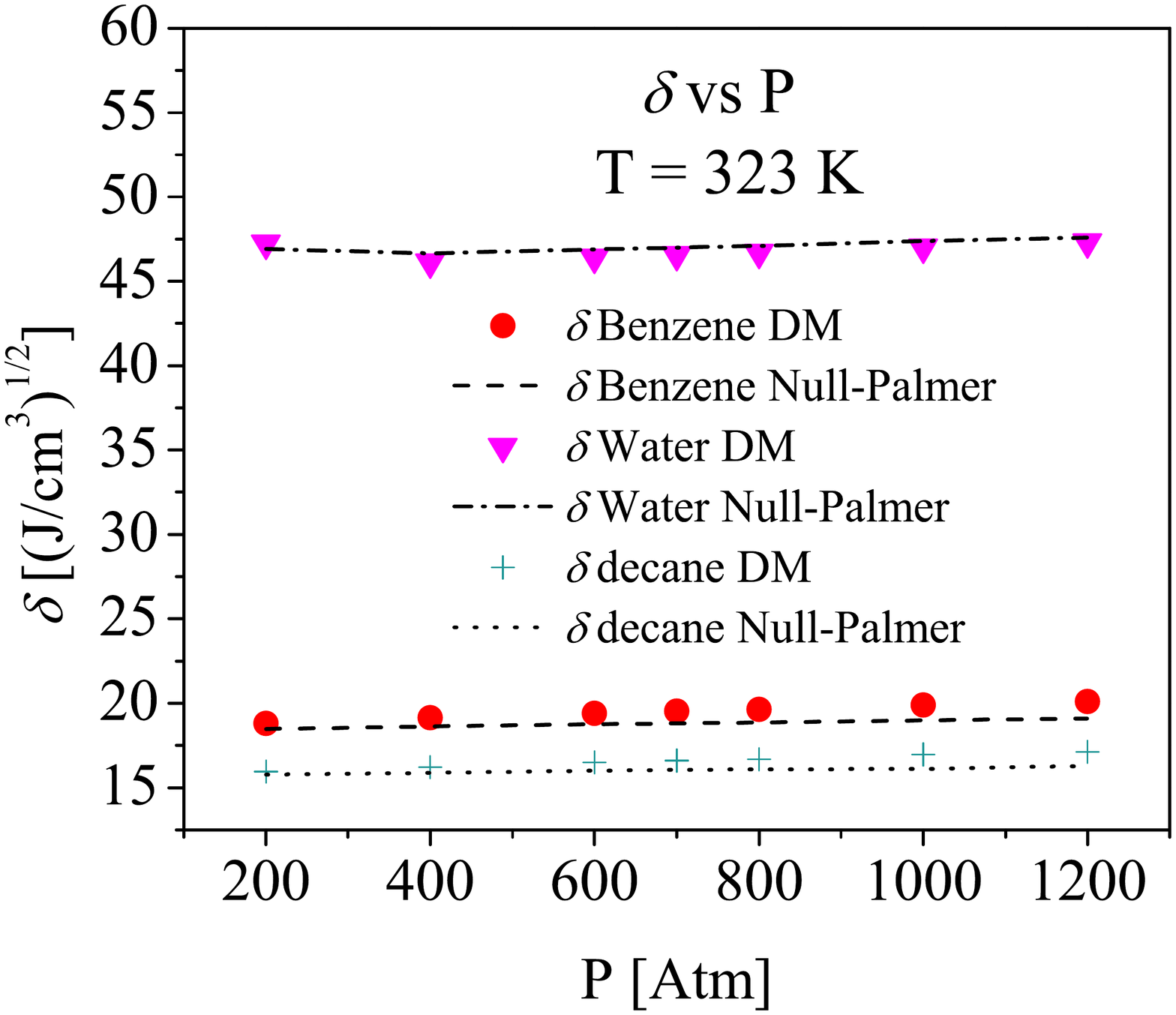}}\\
\begin{center}
\scalebox{0.25}{\includegraphics{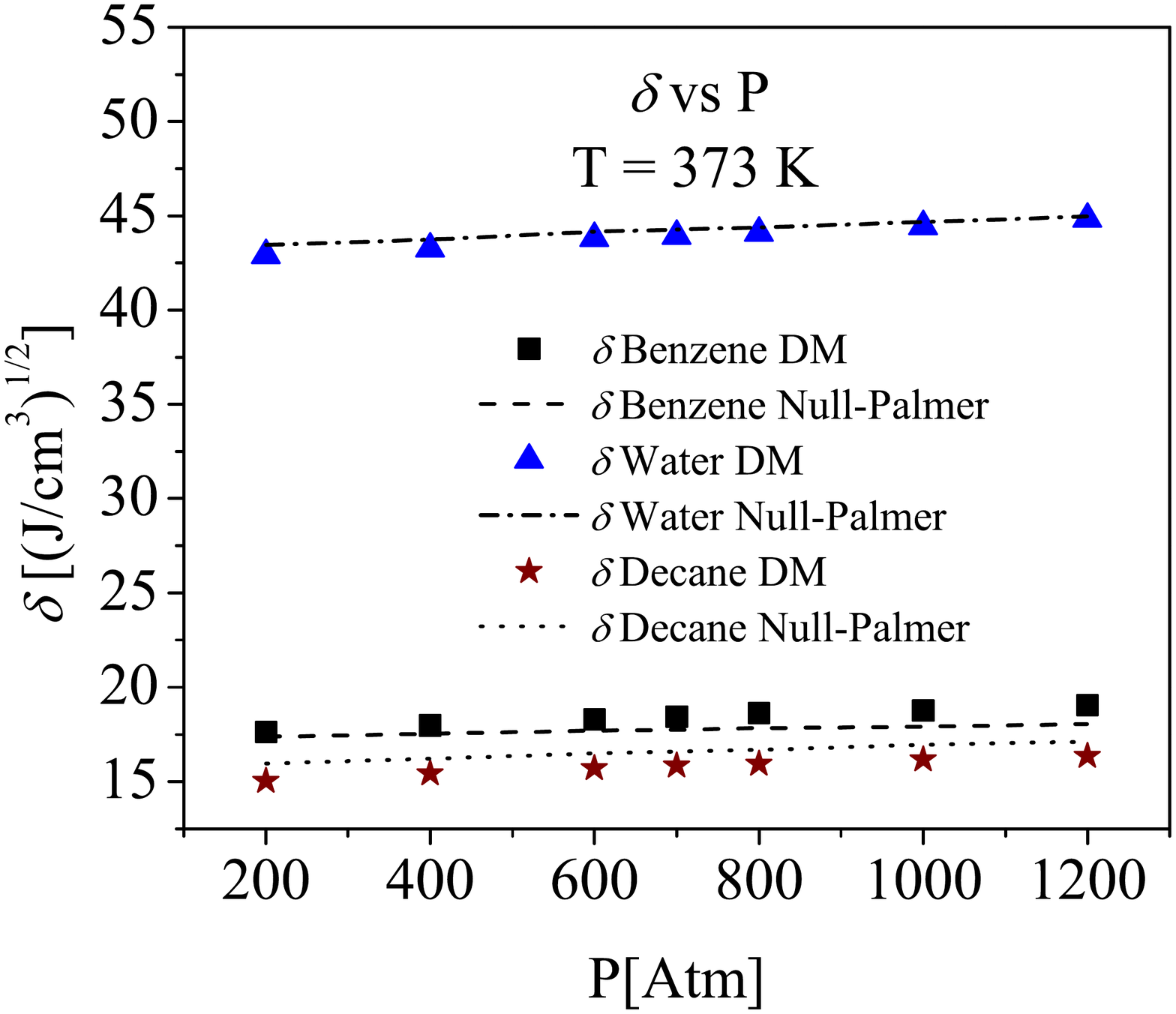}}
\end{center}
\caption{Solubility parameter $\delta[(J/cm ^3)^{1/2}]$ at different $T$ and $P$ for water, n-decane and benzene obtained from Table~\ref{table1} by MD, and compared with the experimental Null-Palmer equation \cite{RefNull}. (See Section~\ref{sec4A} for details).}
\label{psvsP}
\end{figure*}

The results for the solubility parameter $\delta$ and molar volume $V_m$ are presented in Table \ref{table1}. Figure \ref{psvsP} shows our results for water, benzene and n-decane compared to the theoretical prediction using the Null-Palmer equation \eqref{deltanull} and the molar volume obtained by MD simulations. The B, C Antoine's constants were taken from \cite{RefAntoine, antoinedecane} and summarized in Table \ref{table0}. We can observe, as is expected, that at fixed temperature the solubility parameter increases only slightly with pressure. The agreement is striking, with the benefit that our methodology may be employed for any species.

\begin{table*}[h!]
\caption[]{ Antoine's constants for water, benzene and n-decane, taken from \cite{RefAntoine, antoinedecane} }
\begin{center}
\begin{tabular}{ccccccccc}
\hline \hline
 $B_{water}$ & $C_{water}$ & $B_{benzene}$ & $C_{benzene}$ & $B_{n-decane}$ & $C_{n-decane}$\\
\hline
    &  &  T = 298 K & &   & \\
\hline
1730.63	& 233.426 &1196.76	& 219.161&1454.702	&189.265\\
\hline
   &  &  T = 323 K & &   & \\
\hline
1730.63	& 233.426 &1196.76	& 219.161 &1454.702&	189.265\\

\hline
   &  &  T = 373 K & &   & \\
\hline
1810.94&	244.485&1415.8	&248.028&1454.702	&189.265\\

\hline \hline
\end{tabular}
\end{center}
\label{table0}
\end{table*}

\begin{table*}[h!]
\caption[]{Solubilty parameter $\delta$ and molar volume $V_m$ for water, benzene and n-decane  at different T and P obtained by MD simulations, as described in Section~\ref{sec4A}.}
\begin{center}
\begin{tabular}{ccccccccc}
\hline \hline
P  & $\delta_{water}$ & $\delta_{benzene}$ & $\delta_{n-decane}$ & $V_{water}$ & $V_{benzene}$ & $V_{n-decane}$\\
  $\text{[Atm]}$ & $\text{[(J/cm}^3)^{1/2}\text{]}$ & $\text{[(J/cm}^3)^{1/2}\text{]}$ & $\text{[(J/cm}^3)^{1/2}\text{]}$ & $\text{[cm}^3\text{/mol]}$ & $\text{[cm}^3\text{/mol]}$ & $\text{[cm}^3\text{/mol]}$\\
\hline
 &   &  &  T = 298 K & &   & \\
\hline
200 & 47.126  &19.473  & 16.386 & 18.433 & 86.642  & 190.149\\
400 & 47.508 & 19.798 & 16.657 & 18.164 & 85.298 & 187.062 \\
600 & 47.818 & 19.935 & 16.824 & 17.972 & 84.555 & 185.026\\
700 & 47.971  & 20.068 & 16.988 & 17.888 & 83.993 & 183.551\\
800 & 48.131  & 20.300 & 17.080 & 17.786 & 83.208 & 182.566\\
1000 & 48.353 & 20.367 & 17.365 & 17.639 & 82.718 & 179.745\\
1200 & 48.551 & 20.585 & 17.472 & 17.501 & 81.833 & 178.503\\

\hline
 &   &  &  T = 323 K & &   & \\
\hline
200 & 47.264  &18.800  & 15.959 & 18.351 & 89.000 & 193.714\\
400 & 46.150 & 19.153 &16.223  & 18.561 & 87.470 & 190.756 \\
600 &46.427  & 19.393 & 16.502 & 18.379 &  86.279 &187.860 \\
700 &  46.587 & 19.522 &16.598  &  18.299 &  85.780 & 186.849\\
800 &46.705   & 19.625 & 16.678 & 18.207  & 85.315  &185.912 \\
1000 &47.020  & 19.907 &16.956  & 18.000  & 84.157  & 185.454\\
1200 & 47.322 & 20.111 & 17.113 & 17.843 & 83.341 & 181.393\\

\hline
 &   &  &  T = 373 K & &   & \\
\hline
200 & 42.905 & 17.632 & 15.026 & 19.904 & 93.571 & 202.964\\
400 &43.243 & 17.984 & 15.417 & 19.655 & 91.818 & 198.439\\
600 &43.802 & 18.308 & 15.702 & 19.272 & 90.176 & 201.962\\
700 &43.931 & 18.445 & 15.848 & 19.189 & 89.600 & 193.310\\
800 &44.075 & 18.623 & 15.933 & 19.080 & 88.745 & 192.178\\
1000 &44.429 & 18.761 & 16.163 & 18.848 & 87.983 & 189.825\\
1200 &44.835 & 19.051 & 16.358 & 18.597 & 86.695 & 187.504\\
\hline \hline
\end{tabular}
\end{center}
\label{table1}
\end{table*}

\subsection{Interfacial tension of benzene/water and n-decane/water system at different P and T  by DPD simulations. }\label{NR.Lambda}

Using the  values for $V_i(T,P)$, $\delta_i(T,P)$ and $\delta_j(T,P)$ obtained (see Table \ref{table1}) and equations \eqref{aijPT} and \eqref{chiPT}  we calculated the values of the interaction $a_{ij}(T,P)$ and Flory-Huggins $\chi_{ij}(T,P)$ parameters these are summarized  in Table \ref{table2}. The effect of temperature and pressure was considered directly in the conservative force. 
 With this information the DPD simulations were carried out. We took like-like interaction parameters to be equal: $a_{ii}(T,P) = a_{jj}(T,P) = 25$ (see equation \eqref{aii}. Dimensionless number density $\rho^* = \rho r_c^3$ and the dimensionless repulsive parameters $a^*_{ij} = a_{ij}r_c/k_BT$ were used. A total average density of  $\rho^* = 3.0$ was taken, and the masses were all set equal to $1$. Values for the constants $\gamma = 4.5$ and $\sigma = 3$ were used in order to maintain the temperature $k_B T = 1$. We used a reduced time step of $\Delta t^* = \Delta t(k_BT/mr_c^2)^{1/2} = 0.03$ and the standard velocity-Verlet algorithm adapted for the velocity-dependent dissipative force of the DPD model. Periodic boundary conditions in all directions were imposed, and the total number of DPD particles was $4500$ in a cubic box with $L^* = 11.4$. We performed  $100$ blocks of simulations with $10^4$ time-steps each and the interfacial tension was calculated by averaging over the last $75$ blocks according to Equation \eqref{tensionsim}.

 We studied two mixtures: benzene/water and n-decane/water at $T = 298, 323, 373$ [K] and $P = 200, 400, 600, 700, 800, 900, 1000, 1200$ [Atm]. The two systems were modeled as 50:50 binary mixtures of DPD particles of benzene:water and n-decane:water respectively.

We consider here that the DPD method is scale-free as originally proposed \cite{RefGroot} and previously demonstrated by Fuchslin et al. \cite{scalingDPD} for equilibrium systems, using the appropriate scaling scheme for the interactions. This means that the parameters used in the simulations must be independent of the level of coarse graining. In our case, as the interfacial phenomena present no conventional interactions for a typical length scale, one water molecule was represented by one DPD particle, i.e., $N_m=1$. We chose to represent benzene by one DPD particle, and n-decane by two DPD particles joined by a spring with constant $k = 0.1$ in order to preserve entropic contributions. Note that with this choice one DPD particle will only partially contain a benzene molecule, and two DPD particles will only partially contain a molecule of n-decane. The map is, however, appropriate since this is the scale at which interfacial interactions between each species and water take place. Furthermore, the solubility parameters $\delta_{benzene}$ and $\delta_{n-decane}$ used are those of the appropriate species, which eliminates extra degrees of freedom that may be spuriously introduced by the chosen mapping since the microscopic information given by $\delta$ translates in DPD into a collective property essentially integrating over the degrees of freedom which no longer appear at the mesoscopic scale.
Note also that $r_c \sim 1/2r_{DPD}$ , so only nearest-neighbor interactions are being considered.

\begin{figure*}[h!]
\scalebox{0.3}{\includegraphics{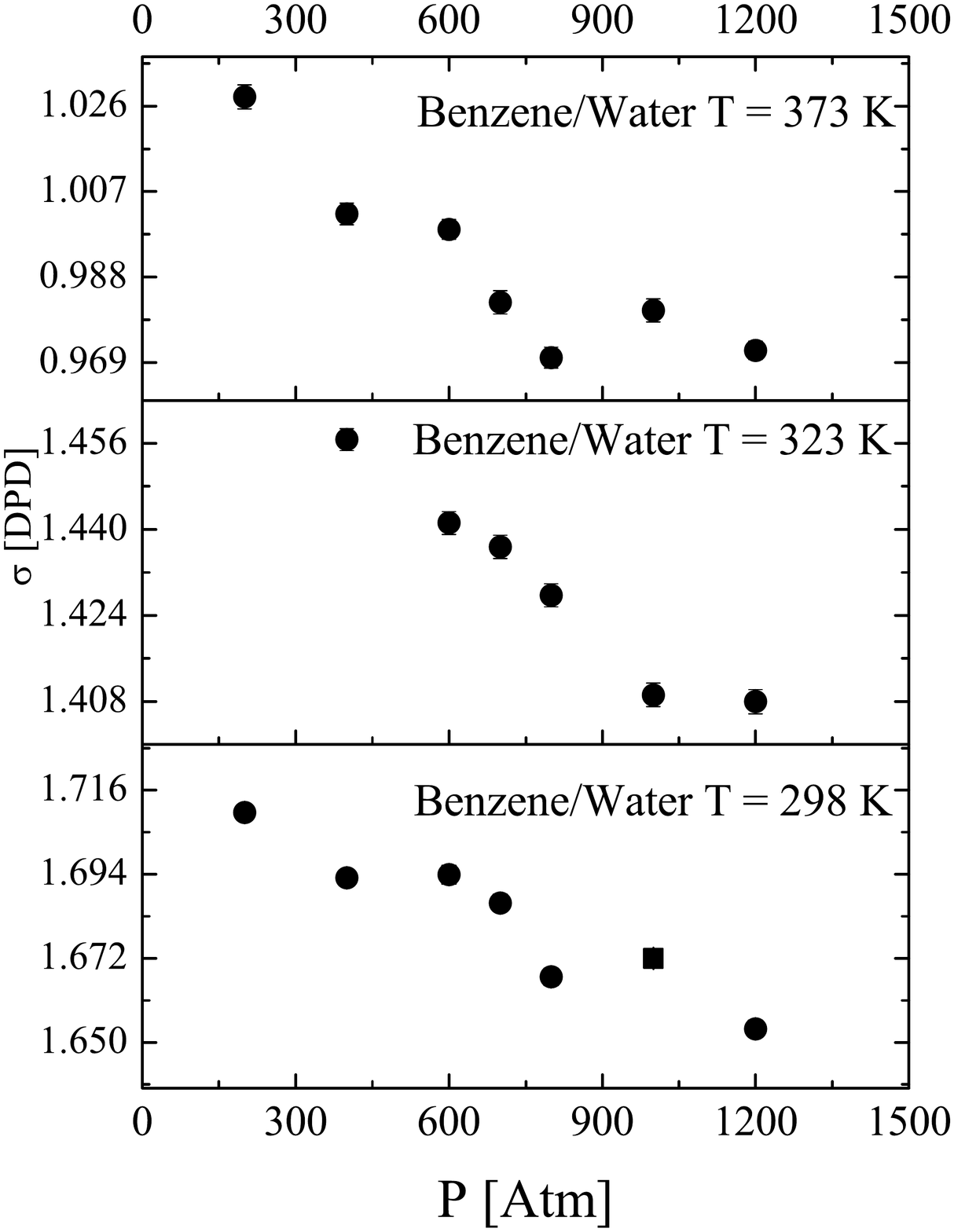}}\quad
\scalebox{0.3}{\includegraphics{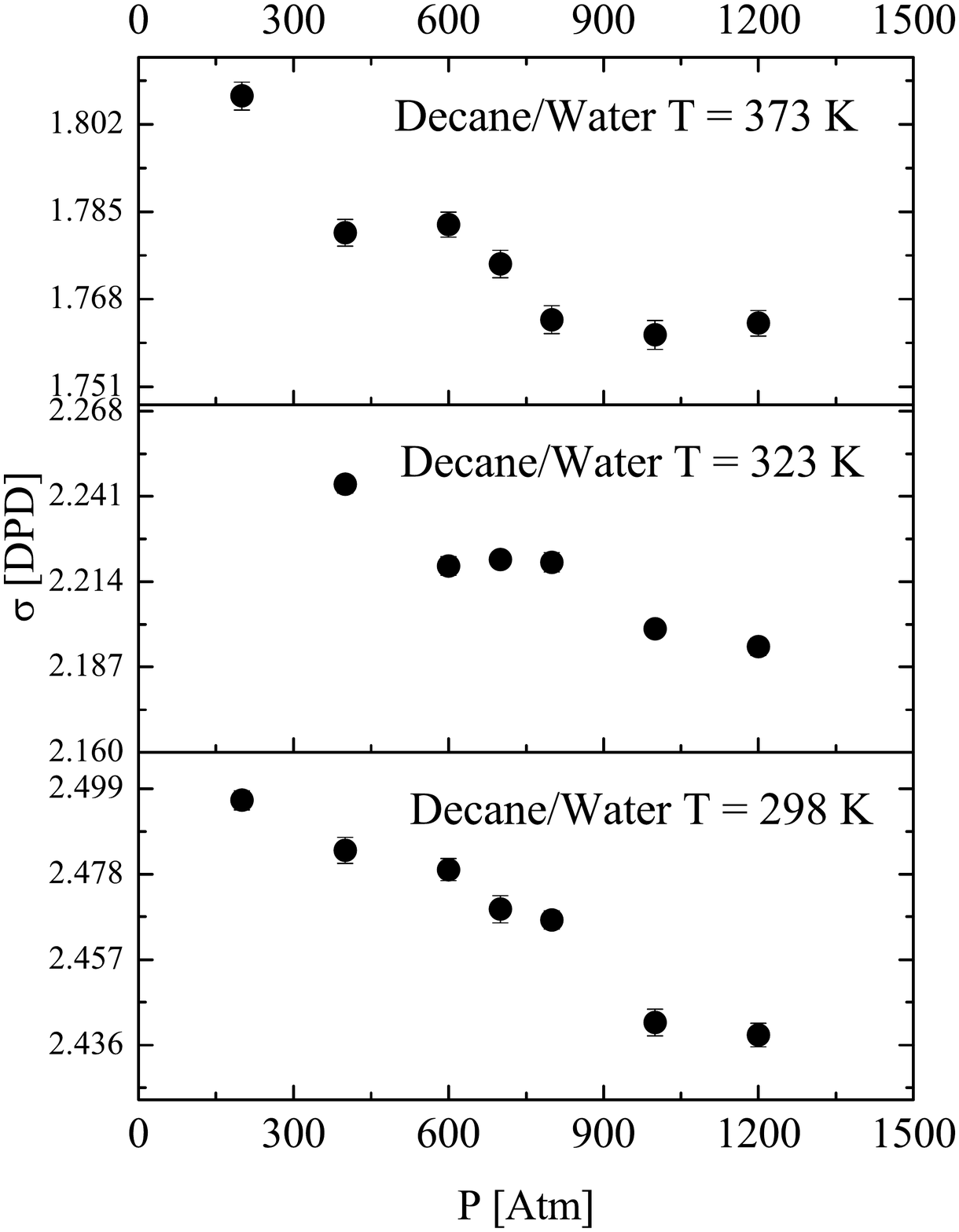}}
\caption{Interfacial tension in DPD units for benzene/water and n-decane/water system at different pressures and temperatures obtained from Table~\ref{table3} by DPD simulations, as described in Section~\ref{NR.Lambda}.}
\label{tensionDPD}
\end{figure*}

\begin{figure*}[h!]
\scalebox{0.27}{\includegraphics{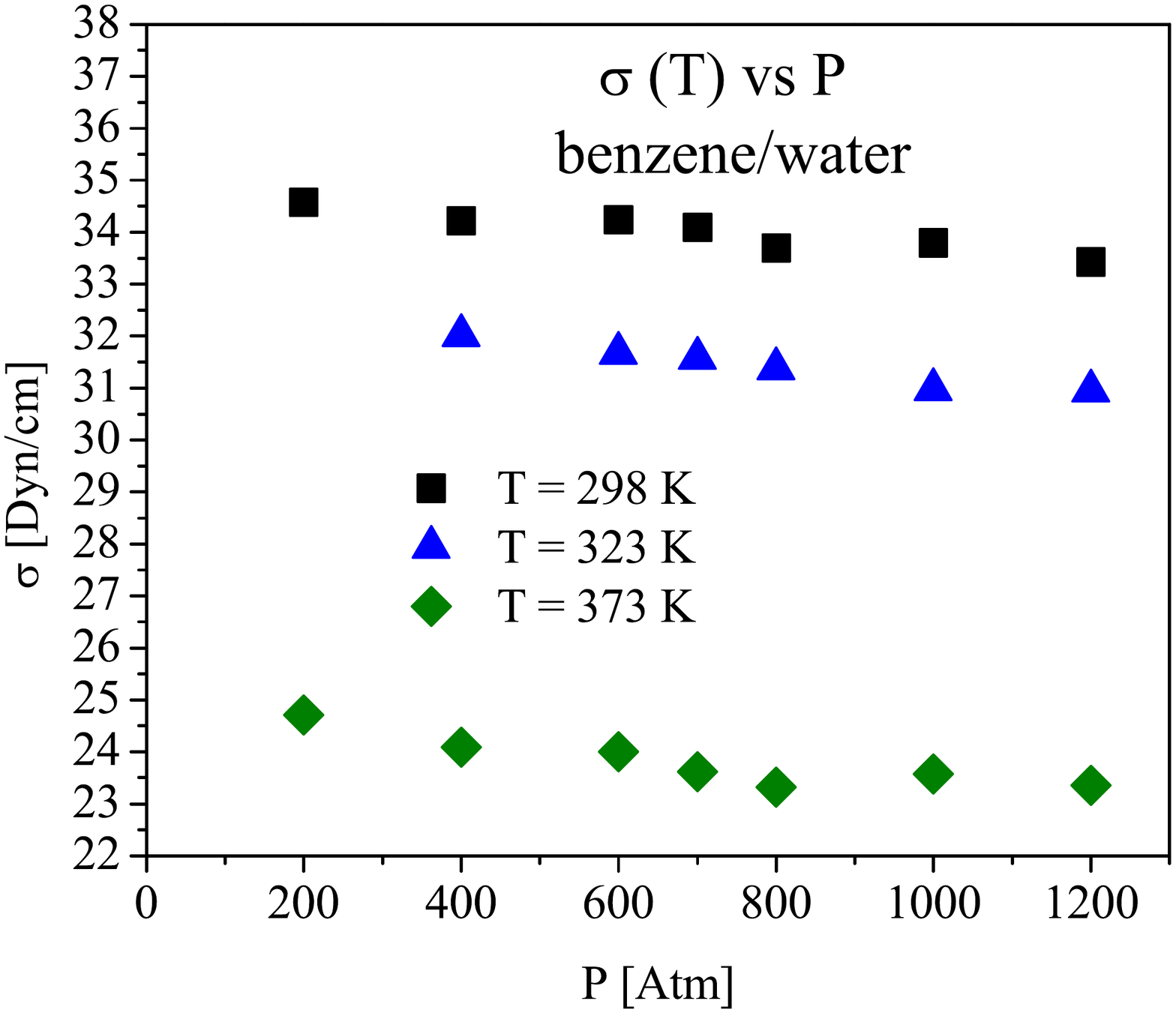}}
\scalebox{0.27}{\includegraphics{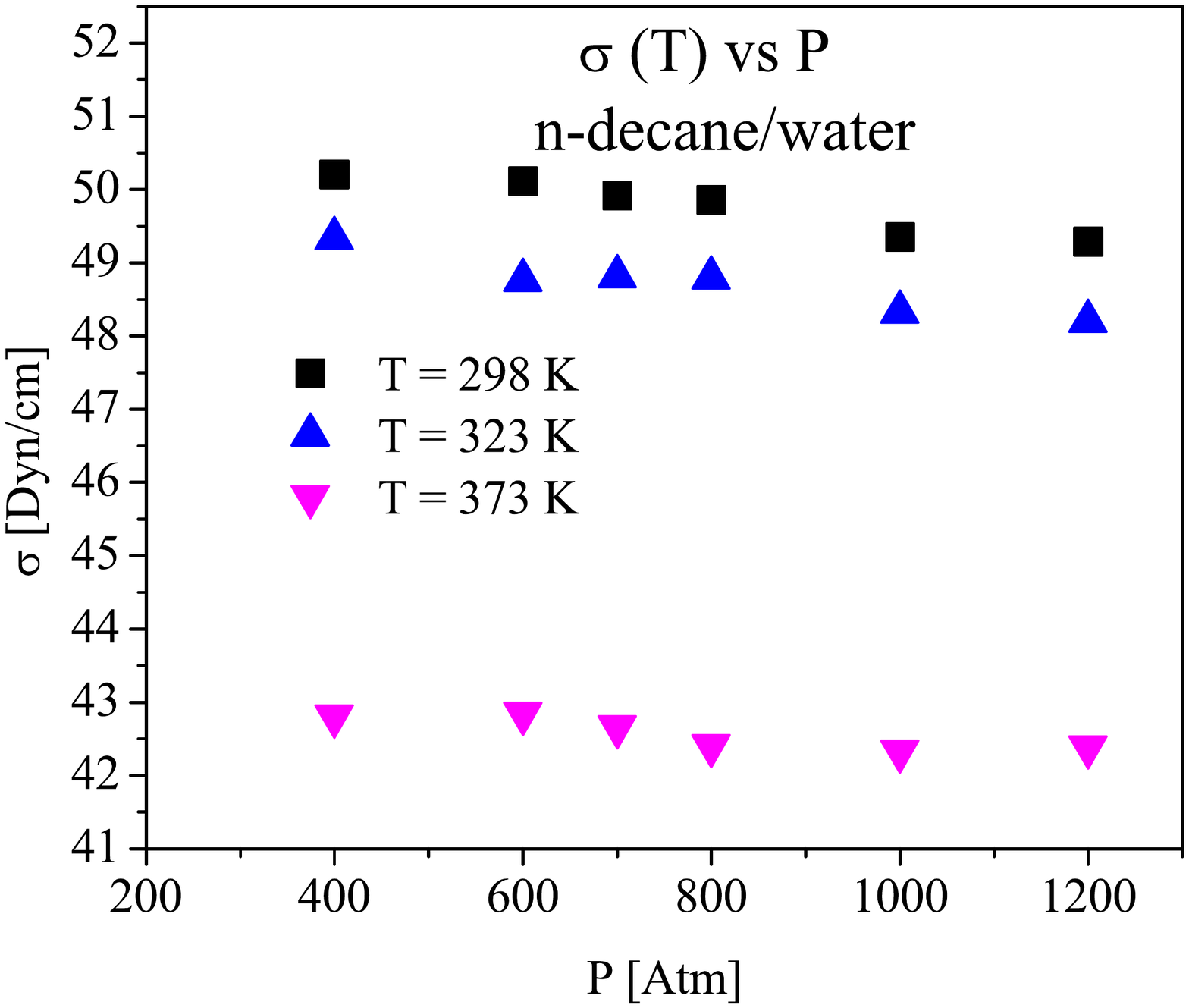}}
\caption{Interfacial tension in physical units for benzene/water and n-decane/water, obtained from Table~\ref{table3} as described in Section~\ref{NR.Lambda} by DPD simulations. Some authors \cite{experimentaltension2, experimentaltension3} report a small and positive pressure coefficient for this systems for pressures lower than 700 atmospheres, while experimental results \cite{experimentaltension1} show a negative pressure coefficient at higher pressures as a consequence of the increased influence in the solubility of the phases. Our results are consistent with the experimental ones.}
\label{tensionesDPDexp}
\end{figure*}

The interfacial tension obtained by DPD simulations using the parameters in Table \ref{table2} is summarized in Table \ref{table3} in DPD units as well as in real units [dyn/cm] obtained using the correspondence between dimensional and adimensional interfacial tension $\sigma_r = (k_BT/r_c^2)\sigma_{DPD}$. Figures \ref{tensionDPD} and \ref{tensionesDPDexp} show these results graphically. The cutoff radius $r_c$ was calculated using equation \eqref{rc}.

\begin{table*}[h!]
\caption[]{Calculated values of $\chi_{ij}(T,P)$ and $a_{ij}(T,P)$ for benzene/water and n-decane/water mixtures, from the values of $V_i(T,P),\ \delta_i(T,P)$ in Table~\ref{table1}, and from eqs.(11,12), using MD. }
\begin{center}
\begin{tabular}{ccccccccc}
\hline \hline
& & &Benzene/Water system& & &\\
\hline
 &T = 298 K& &T = 323 K& &T = 373 K&\\
\hline
Pressure[Atm]&$\chi_{bw}$&$a_{bw}$&$\chi_{bw}$&$a_{bw}$&$\chi_{bw}$&$a_{bw}$\\
\hline
200 & 5.689 & 44.644 & 5.536 & 44.110 & 4.099 & 39.086\\
400 & 5.629 & 44.435 & 5.037 & 42.365 & 4.044 & 38.891\\
600 & 5.639 & 44.470 & 5.001 & 42.240 & 4.039 & 38.875\\
700 & 5.621 & 44.406 & 4.991 & 42.204 & 4.019 & 38.805\\
800 & 5.560 & 44.193 & 4.972 & 42.136 & 3.985 & 38.688\\
1000 & 5.576 & 44.249 & 4.927 & 41.979 & 4.004 & 38.753\\
1200 & 5.524 & 44.068 & 4.920 & 41.954 & 3.987 & 38.692\\
\hline \hline
&&&n-Decane/Water system&&&\\
\hline
&T = 298 K&&T = 323 K&&T = 373 K\\
\hline

Pressure[Atm]&$\chi_{dw}$&$a_{dw}$&$\chi_{dw}$&$a_{dw}$&$\chi_{dw}$&$a_{dw}$\\
\hline

200 & 7.030 & 49.333 & 6.696 & 48.167 & 4.989 & 42.195\\
400 & 6.977 & 49.148 & 6.190 & 46.396 & 4.907 & 41.911\\
600 & 6.968 & 49.116 & 6.128 & 46.181 & 4.907 & 41.911\\
700 & 6.931 & 48.985 & 6.128 & 46.178 & 4.880 & 41.815\\
800 & 6.921 & 48.952 & 6.113 & 46.126 & 4.872 & 41.789\\
1000 & 6.837 & 48.657 & 6.058 & 45.934 & 4.856 & 41.731\\
1200 & 6.823 & 48.607 & 6.064 & 45.954 & 4.863 & 41.756\\
\hline\hline

\end{tabular}
\end{center}
\label{table2}
\end{table*}

\begin{table*}[h!]
\caption[]{Interfacial tension for benzene/water $\sigma_{b/w}$  and n-decane/water $\sigma_{d/w}$  at different $T$ and $P$, both in DPD and physical units, obtained using the results in Table~\ref{table2} from DPD simulations, as described in Section~\ref{NR.Lambda}}
\begin{center}
\begin{tabular}{ccccccccc}
\hline \hline
P  & $\sigma_{b/w} \text{DPD sim}$ &   $\sigma_{b/w} \text{DPD sim} $  & $\sigma_{d/w} \text{DPD sim}$ &  $\sigma_{d/w} \text{DPD sim}$ \\
  $\text{[Atm]}$ & $\text{[DPD units]}$ & $\text{[Dyn/cm]}$ & $\text{[DPD units]}$ & $\text{[Dyn/cm]}$ \\
\hline
 &     & T = 298 K &   & \\
\hline
200	&	1.710	&	34.570	&	2.496		&	50.459	\\
400	&	1.693	&	34.223	&	2.484     &	50.209\\
600	&	1.694	&	34.240	&	2.479	     &	50.113	\\
700	&	1.687	&	34.090	&	2.469 	&	49.916	\\
800	&	1.667	&	33.700	&	2.467 	&	49.863	\\
1000	&	1.672	&	33.797	&	2.442		&	49.352	\\
1200	&	1.654	&	33.425	&	2.438		& 	49.290\\

\hline
  &  &  T = 323 K &  &    \\
\hline
400	&	1.457		&	31.770	&	2.245	 &	48.957	\\
600	&	1.441		&	31.432 	&	2.219	 &	48.393	\\
700	&	1.437		&	31.335	&	2.222	 &	48.439		\\
800	&	1.428		&	31.138	&	2.220	 &	48.420		\\
1000   &	1.409	     &	30.733	&   2.199 &	47.959		\\
1200   &	1.408		&	30.706	&	2.193	 &	47.836		\\

\hline
 &   &    T = 373 K & &    \\
\hline

200	&	1.028	&	24.714	&	1.808	 &	43.450	\\
400	&	1.002	&	24.089	&	1.781	 & 42.811\\
600	&	0.999	&	24.006	&	1.783	 &	42.849	\\
700	&	0.982	&	23.614	&	1.775	 &	42.666	\\
800	&	0.970	&	23.318	&	1.764	 &	42.406	\\
1000   &	0.981	&	23.572	&	1.761	 &	42.335	\\
1200   &	0.972	&	23.358	&	1.763	 &	42.389 \\

\hline \hline
\end{tabular}
\end{center}
\label{table3}
\end{table*}

\begin{figure*}[h!]
\scalebox{0.25}{\includegraphics{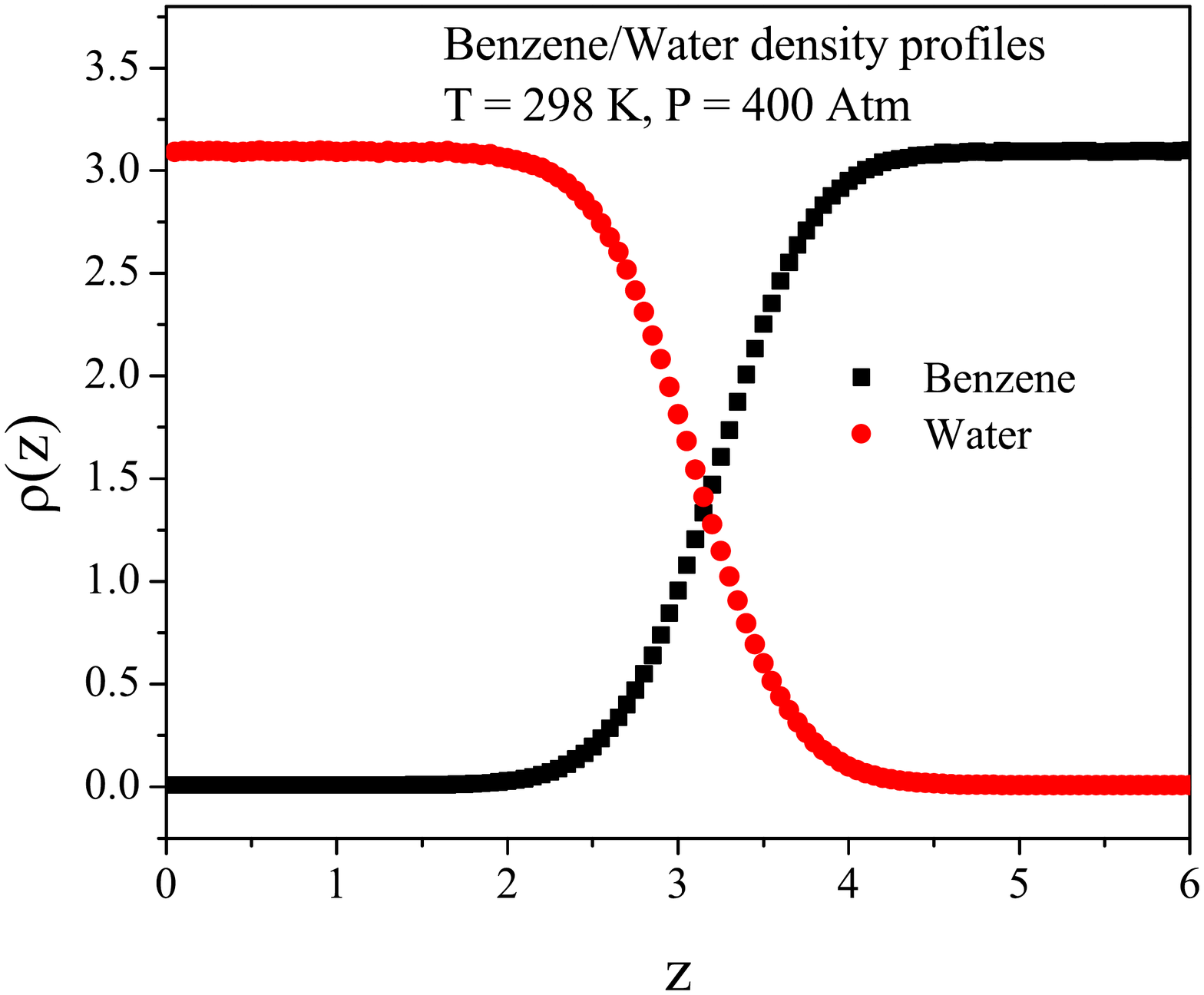}}\quad
\scalebox{0.25}{\includegraphics{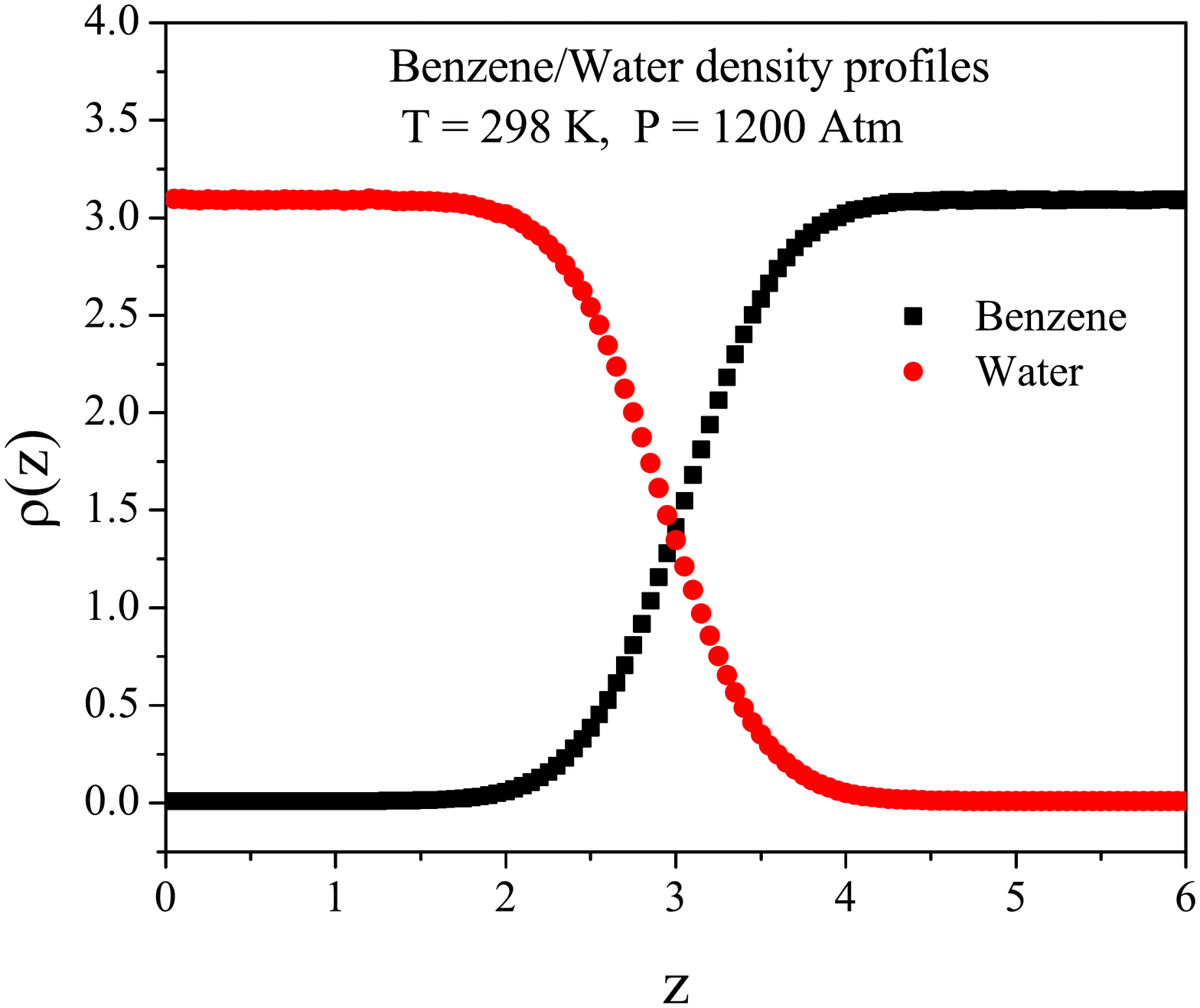}}\\
\scalebox{0.25}{\includegraphics{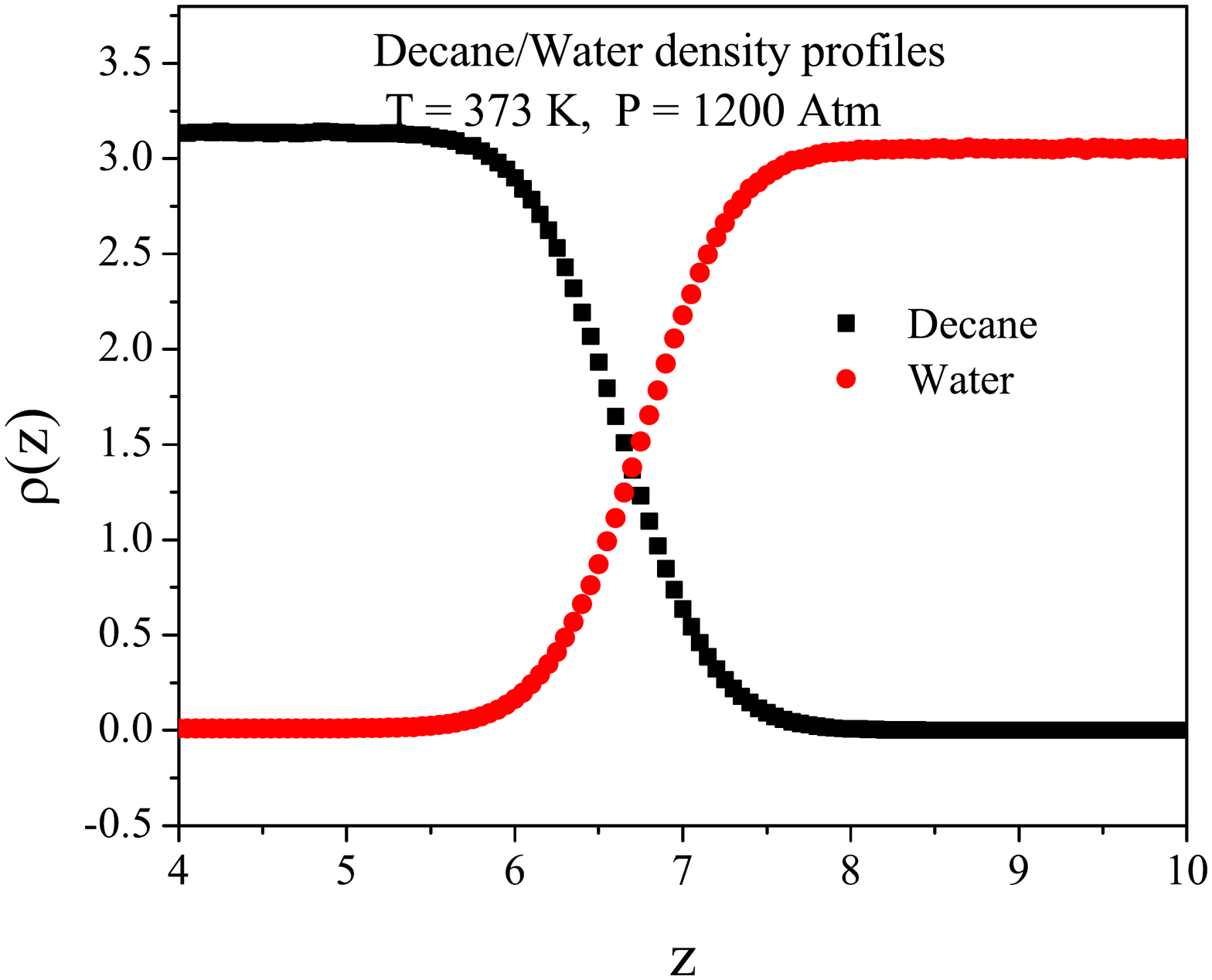}}\quad
\scalebox{0.25}{\includegraphics{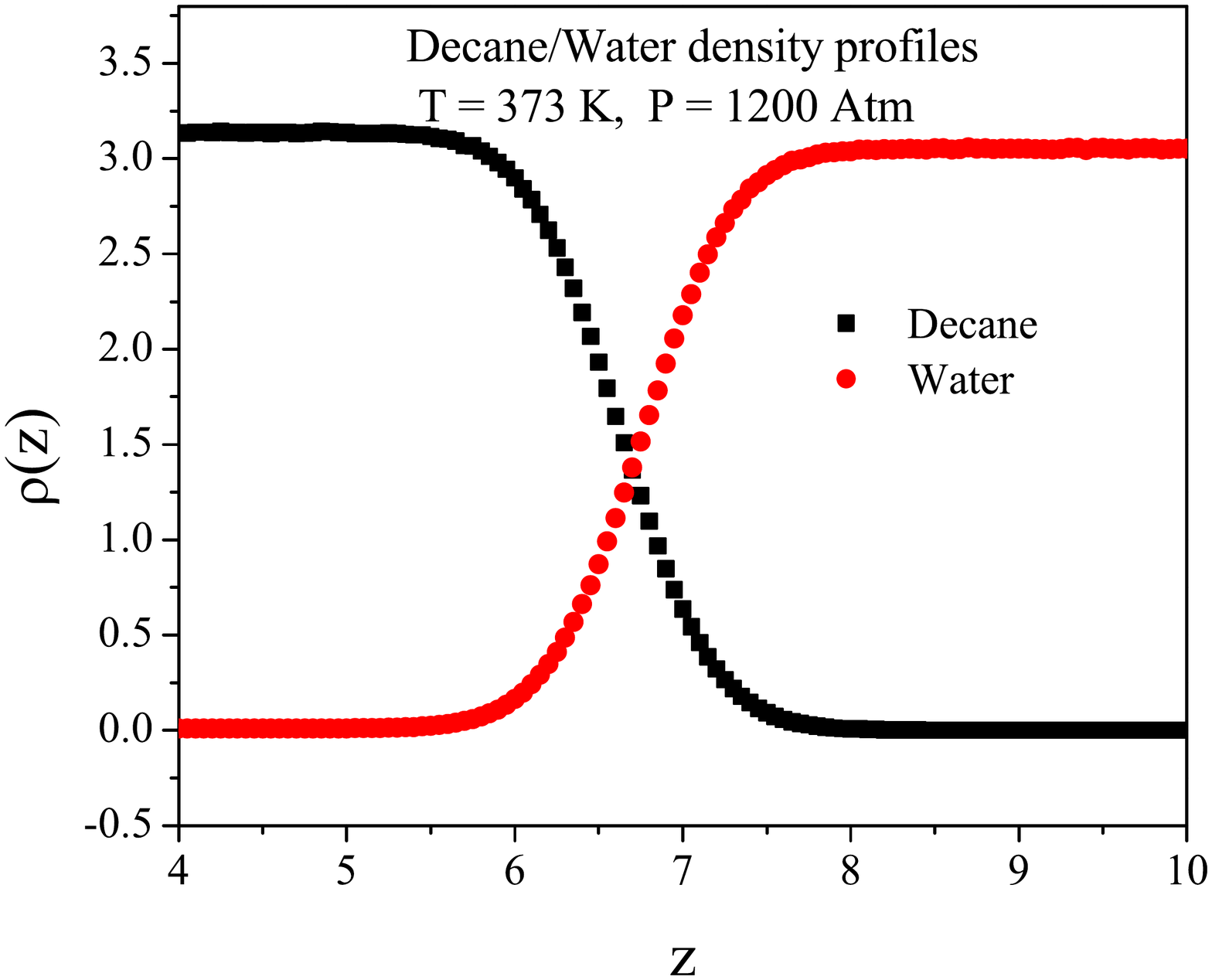}}
\caption{Density profiles for benzene/water and n-decane/water systems at T = 298 K and P = 400 and 1200 Atm, obtained from DPD simulations as described in Section~\ref{sec4A}.}
\label{dpDPD}
\end{figure*}

Even though the maximum experimental change in interfacial tension is less than 3 [dynes/cm]  over the entire pressure range \cite{experimentaltension2} due to the low mutual miscibility between the liquids, as can be observed in the density profiles shown in figure \ref{dpDPD} our simulations can capture this effect because of the direct dependence of the conservative force with the solubility parameters. Differences between benzene-water and n-decane/water mixtures indicate that there are weaker attractive forces between n-decane and water than between benzene and water, which is evident from the $\chi_{ij}$ values obtained.

\section{Conclusions}
A methodology is presented for modeling the pressure dependence of interfacial tension in binary mixtures using a combination of Molecular Dynamics and Dissipative Particle Dynamics simulations. The cohesive energy density and the solubility parameters are calculated at different temperatures and pressures using Molecular Dynamics simulations and the thermodynamic properties of real systems are reproduced with excellent agreement. The pressure and temperature dependence of the Flory-Huggins and repulsive interaction parameters are also obtained and the experimental behavior of the considered systems is reproduced. The methodology described works in a very large range of parametric conditions, is sensitive to changes of only a few dynes/cm,  and is amenable to applications were experimentation may be difficult or even impossible.

\section{Acknowledgements}

We thank DGTIC-UNAM for computational support.

\end{document}